# Single-molecule DNA Bases Discrimination in Oligonucleotides by Controllable Trapping in Plasmonic Nanoholes

*Jian-An Huang[a], Mansoureh Z. Mousavi[a], Yingqi Zhao[a], Aliaksandr Hubarevich[a], Fatima Omeis[a], Giorgia Giovannini[a], Moritz Schütte[b], Denis Garoli[a,c]\*, Francesco De Angelis[a]\**

[a]Istituto Italiano di Tecnologia, Via Morego 30, 16163 Genova, Italy
[b]Alacris Theranostics GmbH, Max-Planck-Straβe 3, D-12489, Germany
[c]AB ANALITICA s.r.l., Via Svizzera 16, 35127 Padova, Italy
\*Email: Denis.Garoli@iit.it, Francesco.Deangelis@iit.it

**Abstract**: Surface-enhanced Raman spectroscopy (SERS) sensing of DNA sequences by plasmonic nanopores could pave a way to new generation single-molecule sequencing platforms. The SERS discrimination of single DNA bases depends critically on the time that a DNA strand resides within the plasmonic hot spot. However, DNA molecules flow through the nanopores so rapidly that the SERS signals collected are not sufficient for single-molecule analysis. Here, we report an approach to control the time that molecules reside in the hot spot by physically adsorbing them onto a gold nanoparticle and then trapping the single nanoparticle in a plasmonic nanohole. By trapping the nanoparticle for up to minutes, we demonstrate single-molecule SERS detection of all 4 DNA bases as well as discrimination of single nucleobases in a single oligonucleotide. Our method can be extended easily to label-free sensing of single-molecule amino acids and proteins.

# Introduction

Surface-enhanced Raman spectroscopy (SERS) allows label-free detection of single analyte molecules by their narrow fingerprint Raman peaks,[1-5] which is promising for biomedical analysis such as single-molecule sequencing.[6, 7] SERS methods usually employ plasmonic metal nanostructures with a sub-10-nm gap feature that, upon laser excitation, exhibit a strong localized electromagnetic field on their surface, termed "hot spots". When molecules enter the hot spots, their Raman signals are excited and enhanced by the electromagnetic field so much that single-molecule sensitivity can be achieved.[8, 9]

SERS sensing based on a flow-through scheme are desirable for many practical applications including lab-on-a-chip diagnostics.[10] However, some challenges have to be addressed for this technology to evolve. Among them, it would be crucial to have control on the time that the analyte molecule resides in the plasmonic hot spot.[11] For example, DNA strands flow through solid-state nanopores so fast (a few microseconds) that SERS signals collected are not sufficient for single-molecule analysis.[7] Another recent work reported that SERS detection of nucleobases by plasmonic nanoslits could not achieve single-molecule resolution until a collection time of 100 ms was used.[6]

In other words, the fast transport of the analyte molecules represents a current obstacle hindering the development of SERS-based flow-through sensors. Recently, different approaches were developed to integrate electrical, optical and thermal forces to control plasmonic nano-object motion at the nanoscale[12-16] and to slow down molecule transport.[17-19] However, reports that combine these effects to demonstrate single-molecule SERS in the flow-through scheme are still missing. By overcoming this challenge, it may create a revolution of diagnostic devices in, for example, DNA sequencing that reached market level years ago utilising solid-state nanopores.[20-22] In comparison to those pioneering nanopore technologies, Raman-based sensors offer even more advantages thanks to much higher discrimination power provided by Raman spectroscopy. For instance, each DNA base can be distinguished in a Raman spectrum of a DNA strand by their own narrow fingerprint Raman peaks.[23-26] In principle, the capability of multiplexing analysis by Raman spectroscopy may pave a way to single-molecule protein sequencing that is challenging but highly desirable in both basic research and diagnostic applications.[27]

In this work, we introduce an electrokinetic approach to control the residence time of biomolecules in a hot spot by trapping a gold nanourchin (AuNU) in a plasmonic nanohole, as shown in Figure 1a. The physical mechanism relies on the balance between the electroosmotic, electrophoretic, and optical forces (Figure 1d). To show the discrimination power of this approach and its potential applications to DNA and protein analysis, we chose nucleotides as molecules of interest. Prior to trapping, the AuNUs were mixed in solution with the nucleotides to allow them to physically adsorb to the surface of the AuNUs (see Methods). Once trapped, the sharp tips of the AuNU will couple with the sidewall of the nanohole creating plasmonic hot spots, which exhibit a greatly enhanced electromagnetic field for SERS detection of the nucleotides already adsorbed on the tips (Figure 1d). By applying electric potentials, we could keep the AuNUs trapped for minutes such that DNA bases could stay in the hot spots long enough to achieve single-molecule analysis. We believe that this approach represents a significant improvement towards nanopore sensors by label-free optical methods.

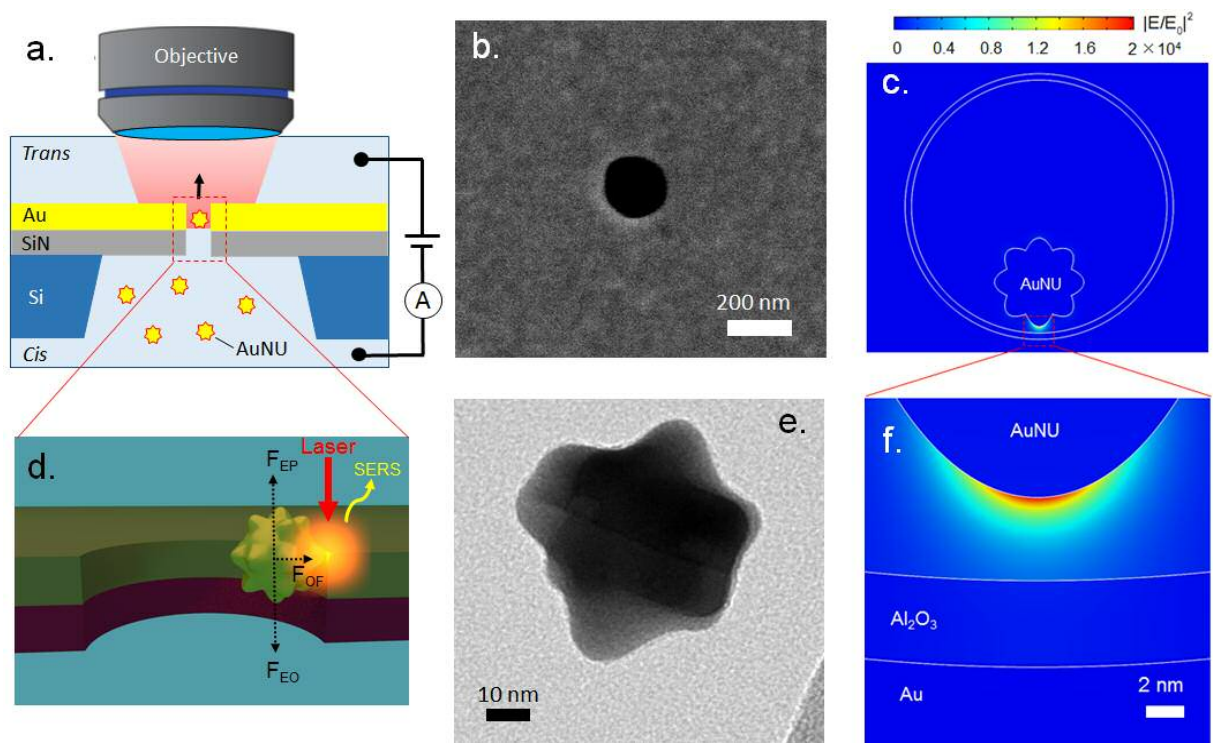

*Figure 1. (a) Schematic of the flow-through setup that allows single AuNUs to flow through and be trapped under applied bias in a gold nanohole with plasmonic resonance upon the laser excitation at 785 nm. The trapping due to a balance between the electrophoretic (EP), electroosmotic (EO), and optical forces (OF) leads to a plasmonic hot spot between the AuNU tip and the nanohole sidewall that allows single-molecule SERS (d). (b) SEM image of the gold nanohole, (e) TEM image of the AuNUs. (c) Simulated electromagnetic field distributions of the AuNU coupled with the nanopore. The color bar represents the enhancement of the electromagnetic field intensity. (f) Magnified field distribution at one tip of the AuNU.*

## Results

**Plasmonic field induced by coupling of a gold nanohole with a AuNU.**

The gold nanohole array with a hole size of 200 nm (Figure 1b) was fabricated by deposition of a 100 nm thick gold film onto a supporting 100 nm thick $Si_3N_4$ membrane before FIB milling of the nanohole array. This nanohole diameter was chosen as it exhibited plasmonic resonance at around the 785 nm laser wavelength (Supplementary Figure S1). Once the 50 nm AuNU (Figure 1e) with adsorbed analyte molecules on its surface flows near the nanohole sidewall, upon laser excitation, the charge oscillation of the AuNU tip couples to the charge of the nanohole sidewall. This induces an intense, confined electromagnetic field at the AuNU tip (Figure 1c, f) for SERS sensing.[28-32] The closer the AuNU tip is to the sidewall, the narrower and stronger the electromagnetic field is found to be (Supplementary Figure S2). As a result, single-molecule detection can be achieved by controlling the proximity between the AuNU and the nanohole's wall as illustrated in Figure. 1d.

**Controllable trapping for reproducible SERS.**

The confinement of the AuNU near the nanohole wall is due to controllable trapping of single AuNUs by turning on and off an electric bias. Among many AuNUs with multilayer Adenines attached on their surface (A-AuNU) that flowed through the nanohole, only those that were close to the sidewall would be trapped and detected, because they produced strong SERS signals. A time trace of the 730 cm$^{-1}$ band intensity of Adenine (red line in Figure 2a) adsorbed on a AuNU at 6 mW laser power and 1V bias indicates that the A-AuNU was trapped in the nanohole until the bias was turned off. The stable and reproducible SERS spectra (Figure 2b) suggested the trapping of a single A-AuNU rather than more than one A-AuNU. If two A-AuNUs are trapped in the nanohole, more than one hot spots will be generated, which will lead to strong fluctuation of both the peak positions and the baseline (Supplementary Figure S3).[33-35] With the feedback from the Raman spectra, the platform could be extended to automated trapping and high-throughput analysis of single AuNUs of interest with a closed-loop program.[36]

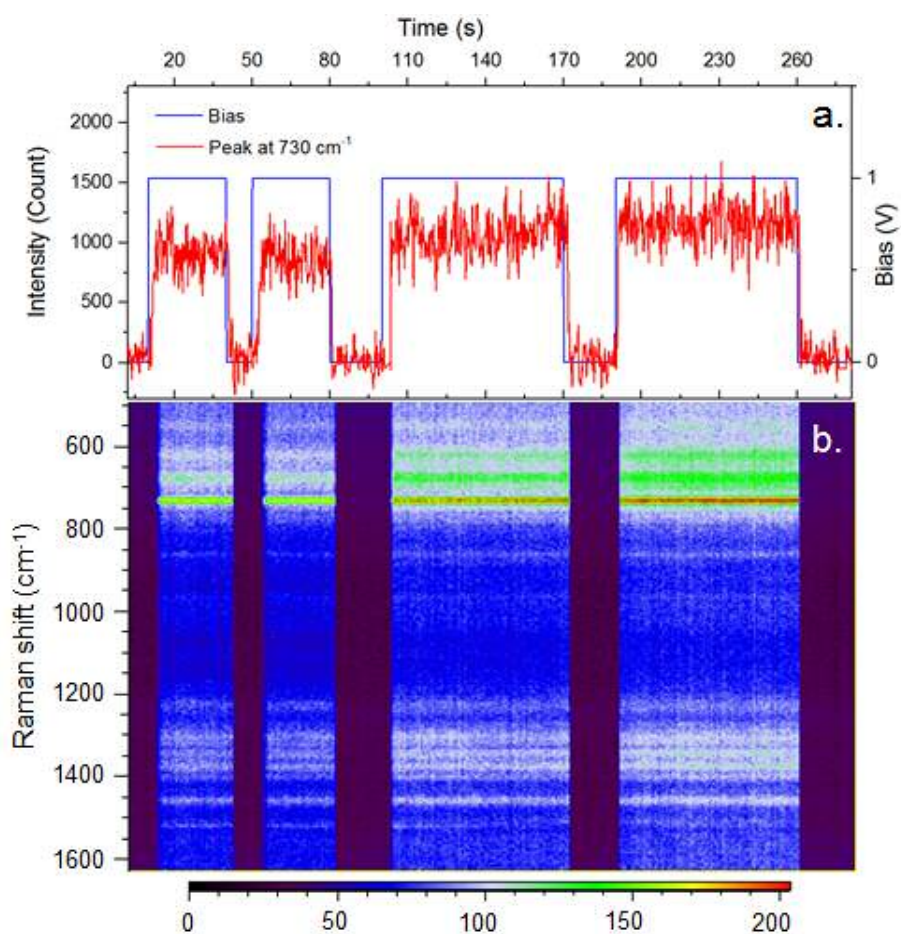

*Figure 2. Reproducible SERS spectra of a trapped AuNU by controllable trapping through turning on and off the bias. (a) Time trace of intensity of the Raman peak at 730 cm$^{-1}$ (red line) of a single A-AuNU trapped in a nanopore by incident laser of 6 mW power and applied bias of 1V (blue line). (b) The contour map of the corresponding stable Raman spectra produced by the trapped A-AuNU. The color bar represents the intensity.*

The stable trapping could last for a few minutes and exhibit reproducible SERS signals with low relative standard deviation (RSD). For example, a RSD as low as ~13% was achieved by 1V bias and 12 mW laser power (Supplementary Figure S4), which was comparable to the RSDs of reproducible solid-state SERS substrates.[10, 37-42] However, high laser power did not always lead to high signal reproducibility, because it also generates more heat, which increases the Brownian motion. Furthermore, since different nucleotide molecules were adsorbed on the AuNUs, electric potential with different amplitudes of 1 – 4 V were used to ensure the stable trapping of AuNUs with different molecules (Supplementary Figure S5). Tuning of both laser power and electrical bias is required to achieve single-molecule SERS.

**Single-molecule SERS of individual DNA bases.**

We applied the bi-analyte SERS technique (BiASERS) for demonstrating single-molecule SERS detection of all 4 DNA bases,[8, 9, 43, 44] in which sub-monolayer combinations of 2 of the 4 DNA bases were attached on the AuNUs (see Methods for details). During collection of more than 1000 Raman spectra, the electromagnetic field could cover either one or both of the two bases and produce the Raman spectra of one or both of the two bases respectively. When the electromagnetic field was sufficiently confined and strengthened, the possibility that the field covers either one of the two bases are higher than covering both, leading to dominance in amount of the Raman spectra of one base over those of both bases. The sub-monolayer nature of the adsorbed bases thus ensured single-molecule detection.

Since the surface areas of the Adenine and Guanine are similar at 1.42 and 1.54 nm$^2$ respectively,[45] we mixed equal moles of both bases with the AuNU solution to allow an average sub-monolayer coverage of around 1 molecule per 2 nm$^2$ on the AuNU surface (AG-AuNU). After SERS detection, we processed the collected SERS data (see Methods for details) to produce a BiASERS histogram of the AG-AuNUs (Figure 3c). The number of spectra for either one of the 2 bases are more than those exhibiting peaks from both DNA bases, proving single-molecule detection.

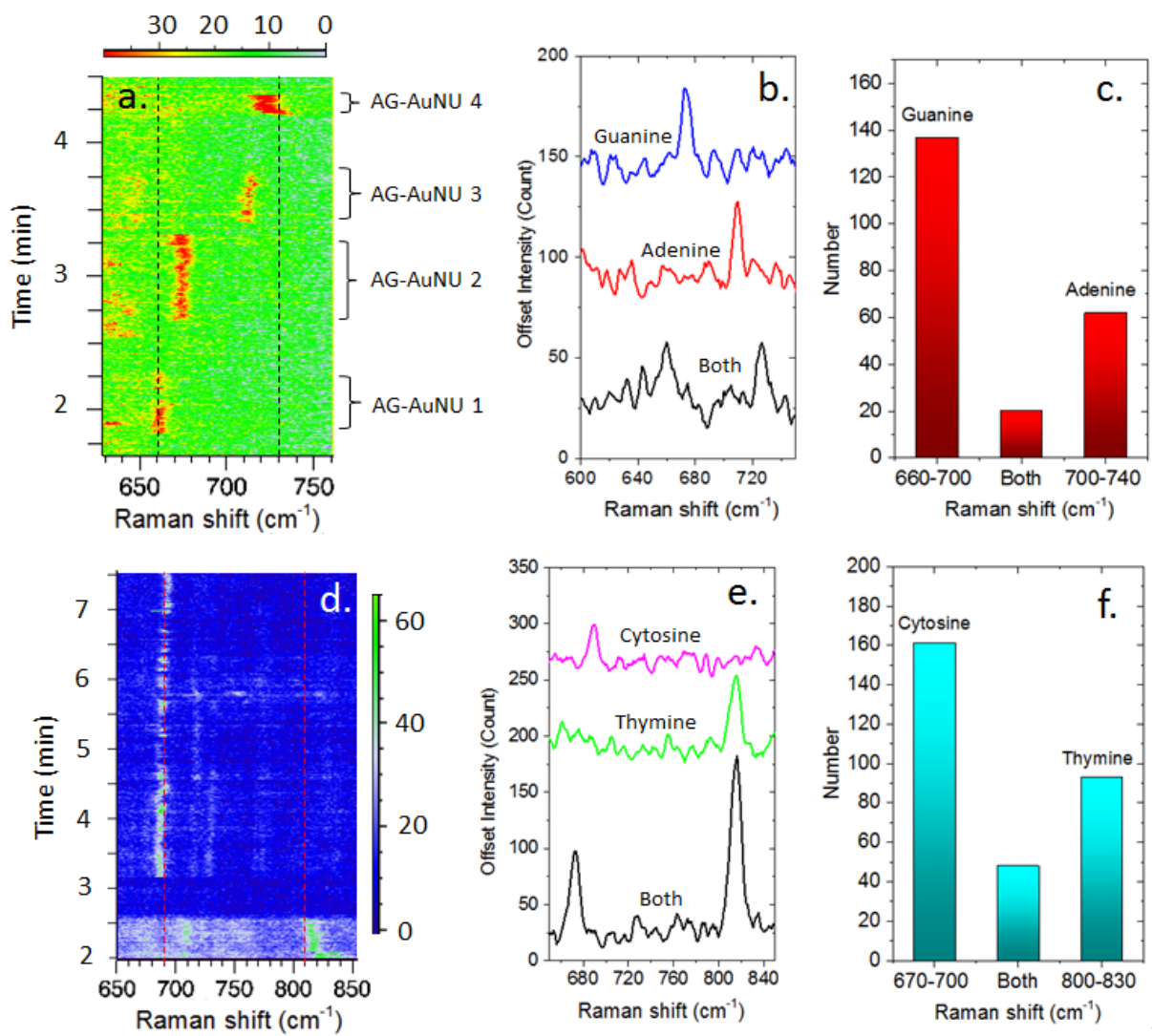

*Figure 3. Single-molecule SERS detection of (a - c) sub-monolayer Guanine and Adenine adsorbed on the AuNUs with molecular surface density of 1 molecule per 2 nm$^2$ under 12 mW laser and 4V bias, and (d - f) sub-monolayer Cytosine and Thymine adsorbed on the AuNUs with molecular surface density of 1 molecule per 7 nm$^2$ under 6 mW laser and 1V bias. (a) A contour map of single-molecule SERS spectra of the AG-AuNUs trapped in a nanohole. The black dotted lines indicate SERS peak positions of multilayer Adenines and Guanines without bias, respectively, from Ref.[46]. The color bars represent the signal-to-baseline intensity of the Raman peaks. (b) Typical SERS spectra of single-molecule events and mixed events of AG-AuNU. (c) Histogram of BiASERS detection of Adenine and Guanine out of about 1200 spectra. (d) A contour map of single-molecule SERS spectra of the CT-AuNUs trapped in a nanohole. The red dotted lines indicate SERS peak positions of multilayer Cytosine and Thymine without bias, respectively, from Ref.[47]. The color bars represent the signal-to-baseline intensity of the Raman peaks. (e) Typical SERS spectra of single-molecule events and mixed events of CT-AuNU. (f) Histogram of BiASERS detection of Adenine and Guanine from about 1500 collected spectra.*

The single-molecule spectra from different AG-AuNUs in Figure 3a exhibit considerable peak shifts of up to 15 cm$^{-1}$ in comparison with SERS spectra of multilayer nucleobase without electrical bias in literatures.[46, 47] For example, the Guanine peaks at around 675 cm$^{-1}$ of the AG-AuNU 2 are red shifted from the multilayer peak at 660 cm$^{-1}$ (black dotted line). This is also seen for the Adenine peaks at around 715 cm$^{-1}$ of the AG-AuNU 3 that are shifted from the multilayer peak at 730 cm$^{-1}$ (black dotted line).

The universal shifting with similar amplitudes for both single-molecule Adenine and Guanine peaks could be due to molecule reorientation on the AuNU surface by the applied electric field that was used to trap the AG-AuNUs.[48-52] Under a static electric field, non-symmetrical molecules will have induced dipole moments and be reoriented by the electric field, exhibiting significant SERS peak shifts.[48, 52] In our case, both Adenine and Guanine molecules are non-symmetrical molecules, because they both have one 6-membered ring and one 5-membered ring. Therefore, they may be easily reoriented by the applied electric potential (4V), leading to strong peak shifts.

Similar single-molecule detection of the Cytosine and Thymine was achieved by mixing equal moles of the two bases with the AuNU solution to form sub-monolayer CT-AuNUs. However, the BiASERS histogram (Figure 3f) was obtained only with lower base concentrations, which on average were equal to 1 molecule per 7 nm$^2$ on the AuNUs. Strong trapping was observed as the Cytosine peak intensity at around 685 cm$^{-1}$ fluctuated from 3.1 to 7.5 min under a 1V bias and a 6 mW laser power. In comparison to the case of AG-AuNUs, the single-molecule SERS peaks of the Cytosine and Thymine shifted with a smaller amplitude (Figure 3d). The small shift amplitude could be due to the small bias (1V) as well as symmetrical molecule structures of Cytosine and Thymine that both have only one 6-membered ring.

**Discrimination of single nucleobases in a DNA**

To demonstrate the discrimination of single nucleobase in a DNA strand, sub-monolayer of oligonucleotides of 5'-CCC CCC CCC A-3'(9C1A), 5'-C AAA AAA AAA-3' (1C9A), and 5'-AAA AAA AAA CTG-3' (9ACTG) were physically adsorbed on the AuNUs, respectively (see Methods), before the AuNUs were driven to the nanoholes for detection. Unlike a normal helix structure in solution, oligonucleotide molecules are adsorbed by nonspecific binding of the nucleobases on the gold surface.[53, 54] The SERS scattering cross-sections of the nonspecific binding nucleobases of a oligonucleotide were found to be A ≈ C > G, T.[55] Due to specific surface selection rules, nucleotides with different conformations on the gold surface will exhibit SERS spectra (Figure 4) different from individual DNA bases.[29, 56-58] Therefore, information of the oligonucleotide conformation can be learnt from the SERS spectra of single nucleobases of the oligonucleotide.[55, 59]

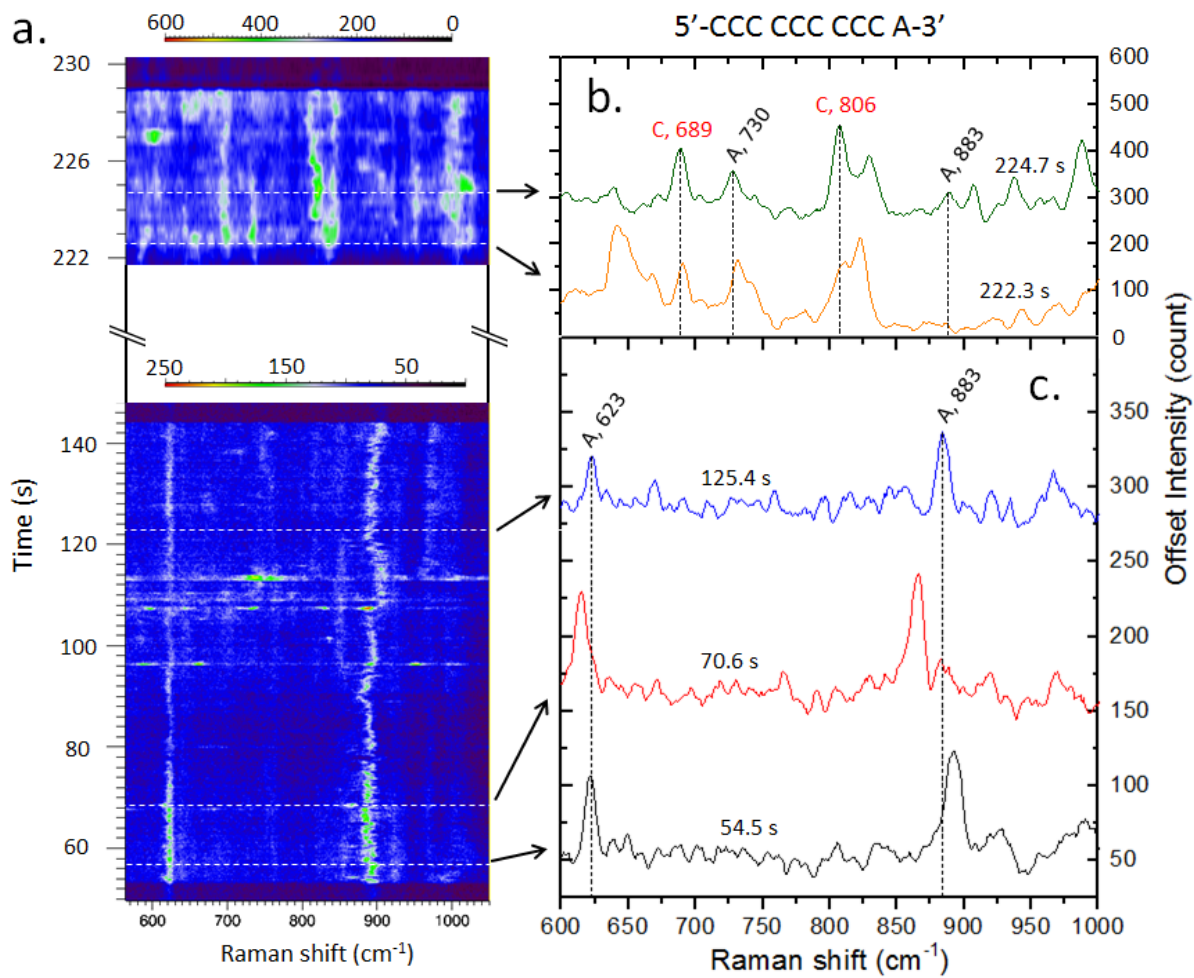

*Figure 4. SERS detection of single nucleobases of the 9C1A oligonucleotide. (a) The contour map of the corresponding Raman spectra extracted from 1400 spectra produced by the trapped 9C1A-AuNUs. The color bars represents the signal-to-baseline intensity of the Raman modes. The white dotted lines indicate (b) the mixed Raman spectra of both A and C, and (c) the single-Adenine spectra, at specific times. The black dotted lines in (b, c) indicate the frequency positions of the Raman peaks unique to A and C, respectively.[46, 47]*

In the single-Adenine spectra of the 9C1A oligonucleotide (Figure 4c), the peaks at around 623 and 883 cm$^{-1}$ were assigned to 6-ring and 5-ring deformations, respectively.[46] Their spectral wandering are a typical signature of single-molecule SERS.[29, 60] However, the strong ring breath mode of Adenine at around 730 cm$^{-1}$ were not present in most single-Adenine spectra (Figure 4a) until mixed spectra of both Adenine and Cytosine were detected at 222.3 s (Figure 4b). The absence of the ring breath mode represents a parallel geometry of the Adenine molecule on the gold surface.[57, 58, 61] Therefore, it suggests that the single Adenine were identified only when the 9C1A molecule was stretched on the AuNU surface presumably by the electroplasmonic forces.[7, 62]

Identification of other single nucleobases were achieved in the cases of 1C9A (long red arrows in Supplementary Figure S6) and 9ACTG (long color arrows in Supplementary Figure S7) oligonucleotides. Their single-base spectra emerged after mixed spectra of other bases were detected, whereas the single-Adenine spectra appeared before the mixed spectra in the case of 9C1A (Figure 5a), which might be due to the strong affinities of Adenines to gold surface.[63] Furthermore, many mixed spectra of both 1C9A and 9ACTG molecules show Adenine Raman peaks except the ring breath mode at around 730 cm$^{-1}$, suggesting again stretched conformations of the oligonucleotides on the AuNUs.

## Discussion

Common BiASERS detections of single molecules usually use dyes and their isotope-edited counterparts. The chosen dyes were resonant with laser illumination to exhibit an enhanced Raman scattering cross-section, and the isotope-edited dyes ensure equal surface adsorption on metal surfaces.[44, 64, 65] However, BiASERS detection of non-resonant single DNA bases by metal nanoparticle dimers either in solutions or on solid-state wafers were not reported. One reason was due to the small Raman scattering cross-section of the DNA bases. For example, the Raman cross-section of Adenine was 10 – 100 times less than the resonant dyes.[64] Besides, the randomly formed nanoparticle dimers could not guarantee a similar field enhancement as well as similar molecule adsorption in the hot spots for all dimers to achieve single-molecule detection events with high probability.

In contrast, our method of trapping single nanoparticles to form hot spots after the nucleobase adsorption ensured that the same bases were excited reproducibly by the same intense electromagnetic field, which led to the collection of many major events of single-molecule spectra. Furthermore, the electromagnetic fields were so confined that single nucleobases were discriminated in the oligonucleotides. Since the multiplexing Raman peaks provide information of all the 4 bases, our method has the potential to monitor DNA conformations on the AuNU tip.

To conclude, by illuminating a plasmonic resonant nanohole, we were able to trap a AuNU in the nanohole on demand by applying an electrical bias. Plasmonic hot spots were formed by coupling the AuNU tip and the nanohole for reproducible SERS detection of nucleotides that were adsorbed on the AuNU before the trapping. The trapping was so controllable that the AuNUs with SERS signals of interest to us could be trapped for minutes. As a result, the nucleotides could stay in the hot spots sufficiently long such that single-molecule detections of all 4 DNA bases and single-nucleobase discrimination in oligonucleotides were demonstrated.

Our method utilises physical adsorption of molecules on the AuNU rather than chemical bonding, which is universal to all molecules. Therefore, our platform can be widely applied to reliable SERS sensing of single molecules such as amino acids and even proteins.

# Methods

**Materials.** Non-functionalized Gold Nano-Urchins (AuNUs) with average particle sizes of 50 nm and concentrations of $3.5 \times 10^{10}$ particles/mL were obtained from Sigma (795380-25ML). The nucleobases adenine (A), cytosine (C), guanine (G), thymine (T) were obtained from Sigma: A (A8626), C (C3506), G (G11950), T (T0376). Phosphate Buffered Saline (806552 Sigma) was used for preparation of the samples and Raman measurements.

**Attachment of multilayer DNA bases on the AuNUs.** The DNA obase molecules were adsorbed on the surface of AuNUs in a condition that still kept the AuNUs stable in electrolyte. To obtain the stable AuNUs-nucleobase, the concentration of AuNUs and the salt concentration were $1.3 \times 10^{10}$ particles/mL of AuNUs and 1.25% of phosphate buffer. The multilayer nucleobase-AuNUs were prepared by pipetting 300 µL of the AuNUs suspension from a stock in an Eppendorf tube, which was then re-suspended in 400 µL of deionised water. Nucleobase solution with final concentration of 125 µM was added in the AuNU solution and allowed to adsorb on the AuNUs with continuous mixing on a shaker at room temperature for 3 hours prior to the Raman measurement. The suspension of the AuNUs with nucleobase were stable in a refrigerator at 4°C for up to two weeks.

**Sub-monolayer attachment of AG and CT on the AuNUs.** For the BiASERS experiments, sub-monolayers of nucleobase were adsorbed on the surface of the AuNUs where lower concentrations of nucleobase solution were used, according to the surface area of nucleobases on gold surface, i.e. A (1.42 nm$^2$), G (1.54 nm$^2$), C (1.27 nm$^2$) and T (1.42 nm$^2$). To obtain the stable AuNUs-nucleobase, the concentration of AuNUs and the salt concentration were $1.3 \times 10^{10}$ particles/mL of AuNUs and 1.25% of phosphate buffer. Then, the average number of nucleobases adsorbed on one AuNU were controlled at 5000 for the BiASERS experiment of AG detection. That is, 50 nM final concentration of each nucleobase solution of Adenine and Guanine were added to the AuNU solution. Similarly, 10 nM of each nucleobase solution of Cytosine and Thymine were added for BiASERS experiments of the CT detection with 1000 nucleobases per AuNU on average. The incubation time to form a sub-mono layer of adsorbed mixture of nucleobase on the AuNUs was 24 hours prior to the Raman measurement.

**Sub-monolayer attachment of oligonucleotides on the AuNUs.** The concentration of oligonucleotides required to achieve a monolayers on the AuNUs surface was determined considering that the surface area of nucleobases on gold surface, i.e. A (1.42 nm$^2$), G (1.54 nm$^2$), C (1.27 nm$^2$) and T (1.42 nm$^2$), and calculating the area occupied by a single oligonucleotide as shown in the table below. The surface area of a single ϕ50 nm AuNUs is calculated as 7850 nm$^2$ by regarding it as a ϕ50 nm gold nanosphere. Knowing the amount of oligonucleotide molecules required to form a monolayer on a single AuNUs, we then calculate the µM of oligonucleotide needed to form a monolayer on AuNUs with a concentration of $1.3 \times 10^{10}$ mL$^{-1}$.

|  | oligonucleotide surface area (nm$^2$) | Number of oligonucleotide /AuNU | µmol of oligonucleotide / $1.3 \times 10^{10}$ mL$^{-1}$AuNUs |
| --- | --- | --- | --- |
| 1C9A | 14.05 | 559 | $9.28\times10^6$ |
| 9C1A | 12.85 | 611 | $1.01\times10^5$ |
| 9ACTG | 17.01 | 461 | $7.66\times10^6$ |

In details, 300 μL of AuNU stock solution were dispersed in 400 μL of 5% PBS pH 5.5 to make $1.3\times10^{10}$ mL$^{-1}$ AuNUs. 100 μL of oligonucleotide solution in the same buffer were added to reach the desired concentration for the monolayer formation (final volume 800 μL): 11.6 nM, 12.68 nM and 9.58 nM respectively for 9A1C, 1A9C and 9ACTG. After vortexing, the samples were left at 4°C for two days allowing the spontaneous absorption of the oligonucleotide molecules on the AuNUs surface. Due to the different conformation and adsorption of the oligonucleotide on the non-uniform AuNUs, the oligonucleotides actually formed sub-monolayer on the AuNUs.

**Measurement of dynamic light scattering and absorbance of the colloid.** DLS experiments were performed using a Malvern Zetasizer and the measurements were evaluated using Zetasizer software. Data are reported as the average of three measurements. Particle diameter, PDI and Zeta potential were used to characterise the colloids suspension before and after functionalisation and to evaluate their stability over time. Unless otherwise mentioned, particles were analysed at a diluted concentration of $1.3\times10^9$ particles/mL in filtered deionised water and in the buffer solution used for their synthesis at 25°C in disposable folded capillary cells (DTS1070). Using aqueous solutions as dispersant at 25°C, 0.8872 cP and 78,5 were used as parameters for density and dielectric constant respectively during the measurements. RI 0,2 and an absorption of 3,32 were used as parameters for the analysis of gold nanomaterials. Cary300 UV-Vis (Varient Aligent) was used for the UV-Vis analysis of the colloidal suspension. The absorbance spectrums were recorded using samples at diluted concentration of $6.5\times10^9$ particles/mL in water or in the buffer used for the synthesis. Absorbance measurements were performed using a 300-800 nm wavelength range in a disposable plastic cuvette (1 mL maximum volume, 1 cm path distance).

**Fabrication of the nanohole devices and PDMS encapsulation.** After sputtering a 2 nm thick titanium and 100 nm thick gold layer on the front side of the $Si_3N_4$ membrane, as well as a 2 nm-thick titanium and 20 nm thick gold layer on its back side, focused ion beam milling (FIB, FEI Helios NanoLab 650 DualBeam) at a voltage of 30 keV and a current from 0.23 to 2.5 nA was used to drill hole arrays in the back of the Ti/Au-coated $Si_3N_4$ sample. The sample was annealed on a hot plate at 200°C in air for 1 hour and allowed to cool naturally. The as-made nanoholes were embedded in a microfluidic chamber made from polydimethylsiloxane (PDMS, Dow Corning SYLGARD 184 silicone elastomer) cured at 65°C for approximately 40 min.

**Simulations.** COMSOL Multiphysics® software was used to perform the modeling of the optical trapping force, electroosmotic and electrophoretic flows (Ref. COMSOL Multiphysics® v. 5.3a. www.comsol.com. COMSOL AB, Stockholm, Sweden). 3D models were constructed to solve the optical trapping, electromagnetic heating and fluid dynamics problems. The optical trapping force was obtained by integrating Maxwell's stress tensor over the particle surface. Perfectly matched layers were used over the whole domain to prevent backscatter from the boundaries. A Gaussian beam was illuminated from the top side. The refractive index and thermal conductivity of gold, silicon and aluminum oxides were taken from the COMSOL library, and the refractive index of the fluid was set to 1.33. The electroosmotic and electrophoretic flows were obtained by solving the Poisson-Nernst-Planck equations.

**Raman measurements.** Raman measurements were obtained by a Renishaw inVia Raman spectrometer with a Nikon 60 × water immersion objective with a 1.0 NA delivering a 785-nm laser beam and an exposure time of 0.1 s. The laser beam was focused to a spot diameter of 1.5 μm with a power varying from 2 to 12 mW.

**Single-molecule data processing.** Spectra were processed using custom python scripts according to Chen *et al.*.[6] Data were selected in a 500-1000 cm$^{-1}$ window and smoothed using a Savitzky-Golay filter. A baseline was fitted to each spectrum using 5th-order polynomial functions and subsequently removed from the spectra. Peaks were detected on the resulting spectra using pythons' signal_find_peaks_cwt function. Final peaks were selected if the peak height exceeds 2 standard deviations of the spectrum.

# Additional Information

**Supplementary information.**

# Acknowledgments

The authors thank Dr. M. Dipalo and Dr. D. Darvill for drawing of schematic figures and editing respectively. The research leading to these results has received funding from the Horizon 2020 Program, FET-Open: PROSEQO, Grant Agreement no. [687089].

# Author Contributions

J.A.H. designed the experiment, carried out Raman experiments, and wrote the manuscript. M.Z.M. and G.G. handled the molecules' attachment on the AuNUs and the AuNU stability. G.G. measured the Zeta potentials and absorbance of the functioned AuNUs. Y.Z. fabricated the nanohole device. A.H. and F.O. did the simulation and investigated the trapping mechanism. J.A.H, A.H. and M.S. analyzed the data. D.G. and F.D.A. supervised the work. All authors discussed the results and contributed to the final manuscript preparation.

# Supporting Information for

# Single-molecule DNA Bases Discrimination in Oligonucleotides by Controllable Trapping in Plasmonic Nanoholes

*Jian-An Huang[a], Mansoureh Z. Mousavi[a], Yingqi Zhao[a], Aliaksandr Hubarevich[a], Fatima Omeis[a], Giorgia Giovannini[a], Moritz Schütte[b], Denis Garoli[a,c]\*, Francesco De Angelis[a]\**

[a]Istituto Italiano di Tecnologia, Via Morego 30, 16163 Genova, Italy
[b]Alacris Theranostics GmbH, Max-Planck- Straβe 3, D-12489 Berlin, Germany
[c]AB ANALITICA s.r.l., Via Svizzera 16, 35127 Padova, Italy
*Email: Denis.Garoli@iit.it, Francesco.Deangelis@iit.it

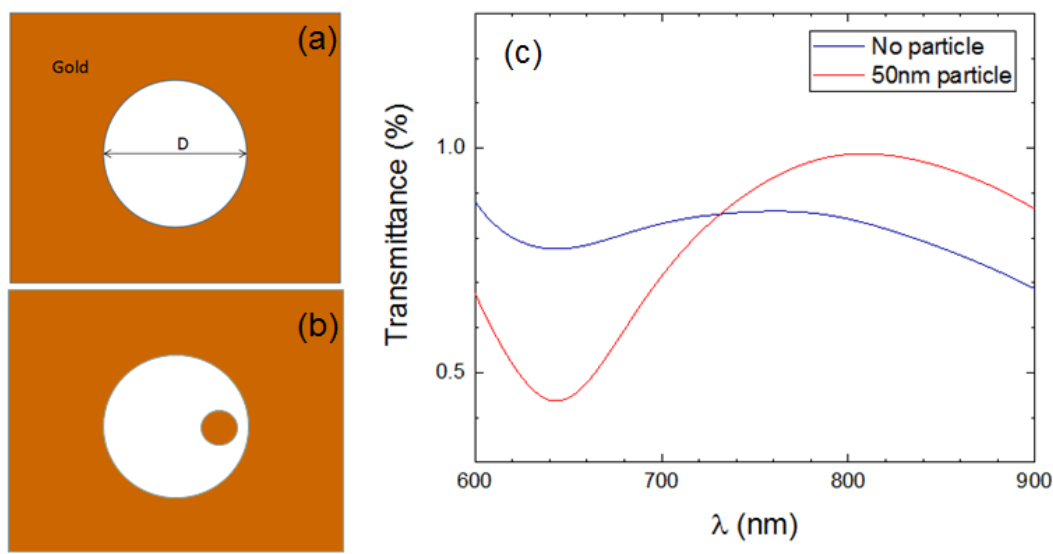

*Figure S1. Simulation model of the gold nanohole (a) and gold nanoparticle in the nanohole (b). (c) Simulated far-field transmission of the ϕ200 nm gold nanohole with and without a ϕ50 nm nanoparticle in it.*

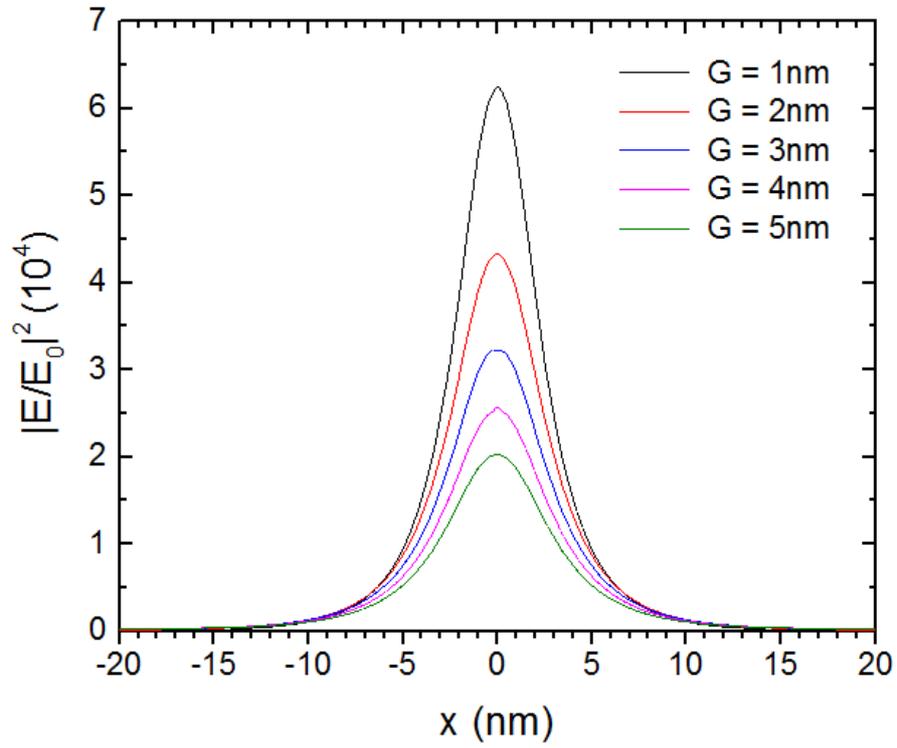

*Figure S2. Simulated field intensity enhancement and distribution on the AuNU tip with different gap distance (G) between the AuNU tip and the alumina layer coated on the nanohole, suggesting that smaller gap leads to stronger and more confined electromagnetic fields.*

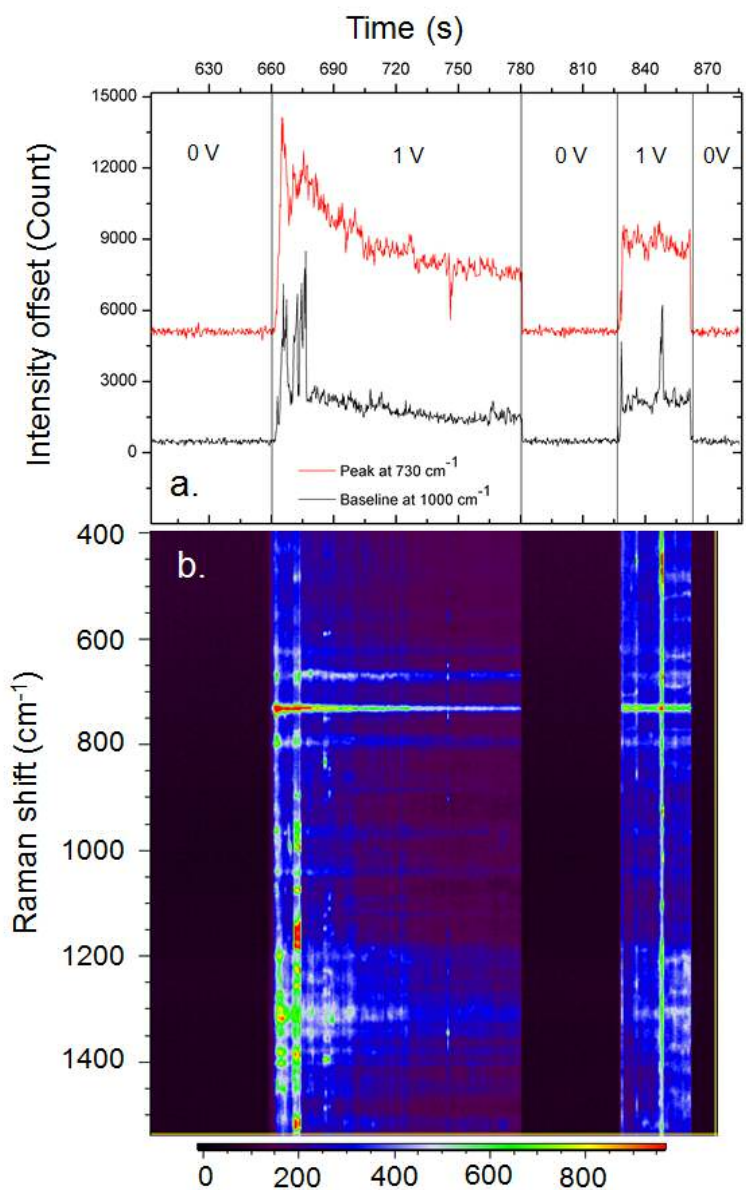

*Figure S3. (a) Time trace of A-AuNU trapped and released in a nanopore by 1V bias and laser power 12 mW. (b) The contour map of the corresponding Raman spectra produced by the trapped A-AuNU. At 660s, trapped AuNUs produced irreproducible SERS signals as well as fluctuating baseline until 680s, which may be because there is >1 AuNUs in the nanohole. After 680s, the baseline became stable, suggesting only one AuNU was left trapped in the nanohole.*

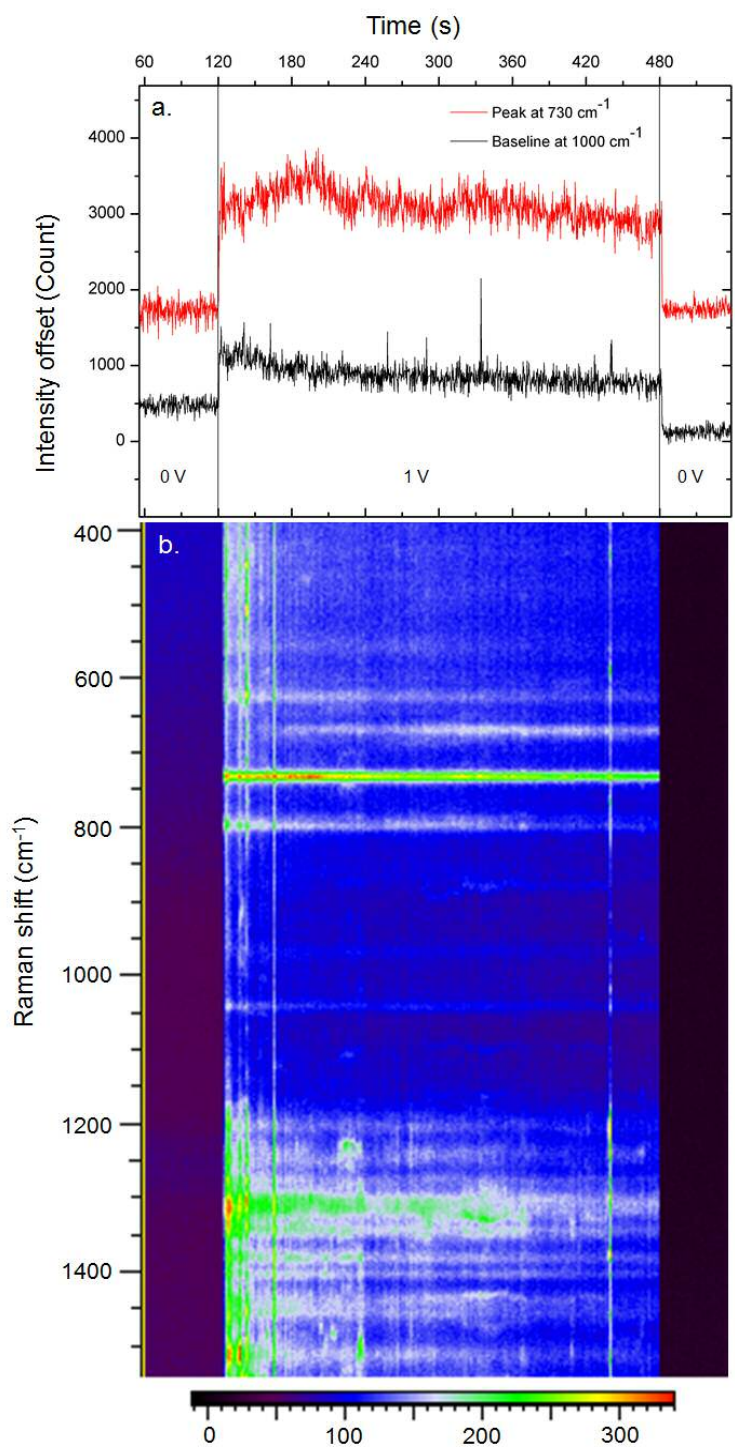

*Figure S4. Reproducible SERS spectra of a trapped AuNU with RSD around 13%. (a) Time trace of Raman peak at 730 cm-1 (red) and baseline at 1000 cm-1 (black) of A-AuNU trapped in a nanopore by 1V bias and laser power 12 mW for 6 minutes until bias was turned off. (b) The contour map of the corresponding stable Raman spectra produced by the trapped A-AuNU.*

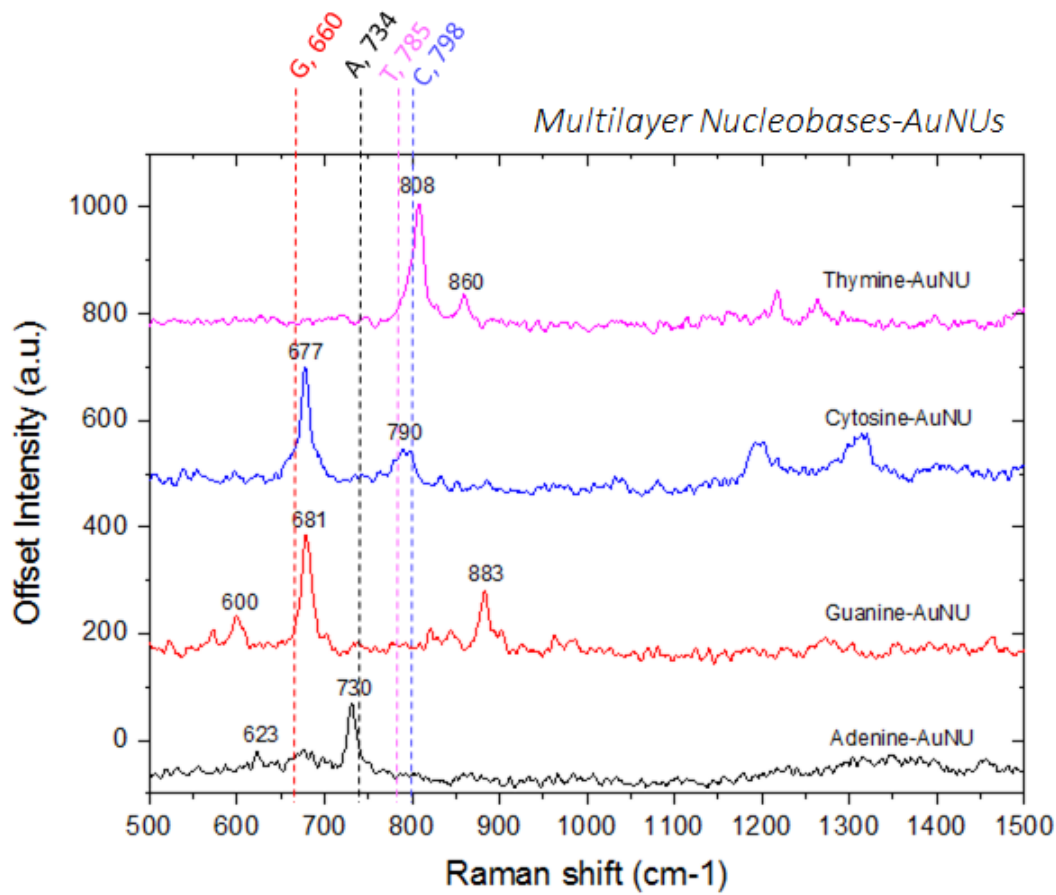

*Figure S5. SERS spectra of multilayer nucleobase-AuNUs trapped in the nanohole under different trapping conditions: A-AuNU (1V, 6 mW), C-AuNU (1V, 12 mW), G-AuNU (2V, 6 mW) and T-AuNU (4V, 6 mW). The colored dotted lines indicate the band positions of the strong ring breathing modes of each multilayer-nucleobases enhanced by silver nanoparticles without DC electric field from reference No.[46]. The fact that the shift of the breathing modes increases with the magnitude of the applied bias suggests molecule reorientation on the AuNU surface by the applied electric field.*

*Table S1. Assignment of the SERS bands of the multilayer nucleobase-AuNU in Figure S5 to specific vibrational modes of the nucleobases.*

| A | C | G | T | Assignment[46, 47, 66] |
|---|---|---|---|---|
| Raman shift (cm$^{-1}$) | | | | |
| 623 | | | | In-plane: 6-ring deformation |
| | 677 | | | Out-of-plane: wag N8-H |
| | | 681 | | In-/Out-of-plane: 6-ring breathing, 5-ring deformation, wagging NH$_2$ |
| 730 | | | | In-plane: ring breathing |
| | 790 | | | In-plane: ring breathing |
| | | | 808 | In-plane: ring breathing |
| | | | 860 | In-plane: ring deformation |
| | | 883 | | In-/Out-of-plane: 5-ring deformation, 6-ring deformation, wagging N9-H, N1-H |

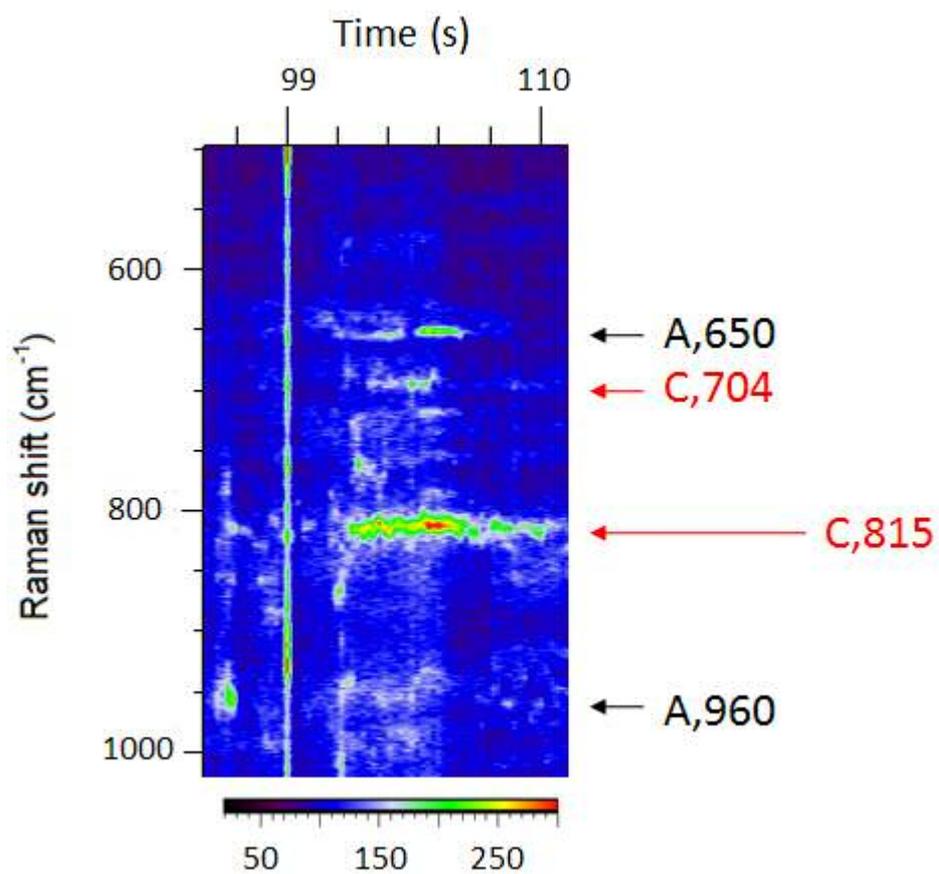

*Figure S6. Contour map of the corresponding SERS spectra of 1C9A oligonucleotides, which is extracted from 1050 spectra produced by the trapped 1C9A-AuNUs. The color bar represents the intensity. The arrows indicate the frequency positions of the Raman peaks unique to either A (black) or C (red).*[46, 47] *Single-Cytosine events occurred between 109 and 112.5 s, as indicated by the long red arrow at C, 815.*

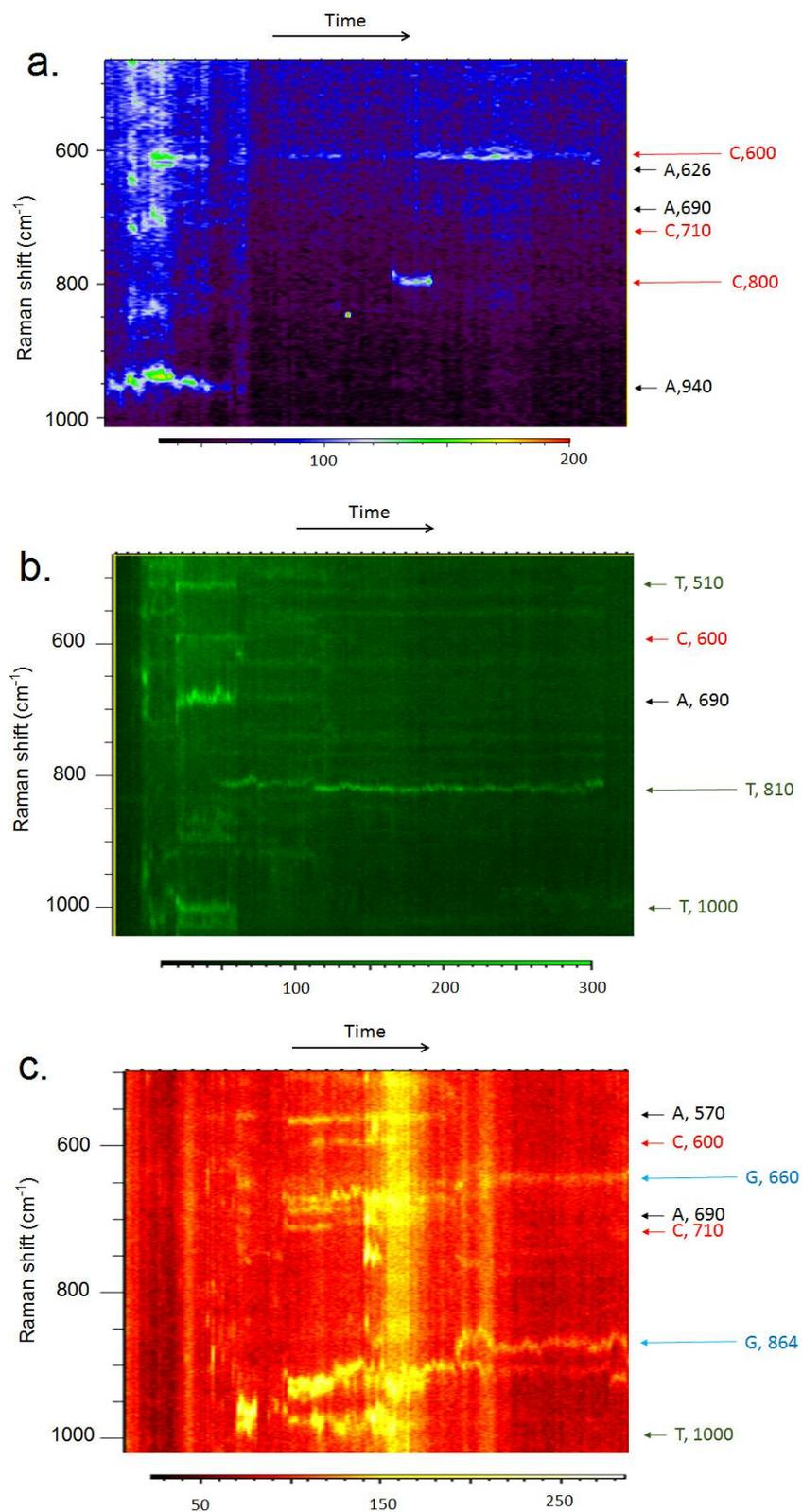

*Figure S7. Contour maps of SERS spectra of different 9ACTG-AuNUs to demonstrate sensing of (a) single Cytosine, (b) single Thymine, and (c) single Guanine, respectively. The color bars represent the peak intensity. The arrows indicate the frequency positions of the Raman peaks unique to either A (black), C (red), T (green) or G (blue).[46, 47] Single-base events were indicated by the long color arrows.*